\documentclass[manuscript]{raa}
\usepackage{graphicx,times}             
\usepackage{natbib}
\usepackage{amsfonts}
\usepackage{txfonts}

\newcommand{\Rmnum}[1]{\expandafter\@slowromancap\romannumeral #1@}
\begin{document}

   \title{2MASS photometry and age estimate of globular clusters in the outer halo of M31
}
   \volnopage{Vol.0 (200x) No.0, 000--000}      
   \setcounter{page}{1}           

   \author{Jun Ma \mailto{}
      \inst{1,2},
      }

   \institute{$^{1}$National Astronomical Observatories, Chinese
Academy of Sciences, Beijing, 100012, P. R. China\\
$^{2}$Key Laboratory of Optical Astronomy, National Astronomical Observatories, Chinese Academy of
Sciences, Beijing, 100012, China\\
             \email{majun@nao.cas.cn}
          }

   \date{Received~~2001 month day; accepted~~2001~~month day}


\abstract{
We present the first photometric results in $J$, $H$, and $K_s$ from 2MASS imaging of 10 classical globular clusters in the far outer regions of M31. Combined with the $V$ and $I$ photometric data from previous literature, we constructed the color-color diagram between $J-K_s$ and $V-I$. By comparing the integrated photometric measurements with evolutionary models, we estimate the ages of these clusters. The results showed that, all of these clusters are older than $3\times 10^9$ yrs, of which 4 are older than 10~Gyrs and the other 6 are in intermediate ages between $3-8$~Gyrs. The masses for these outer halo GCs are from $7.0\times 10^4 M_{\odot}$ to $1.02\times 10^6~M_{\odot}$. We argued that, GC2 and GC3, the ages, metallicities and the distance moduli of which are nearly the same, were accreted from the same satellite galaxy, if they did not form {\it in situ}. The statistical results show that, ages and metallicities for these 10 M31 outer halo GCs do not vary with projected radial position, and the relationship between age and metallicity doest not exit.
   }
   \authorrunning{Jun Ma}            
   \titlerunning{2MASS photometry and age estimate of globular clusters in the outer halo of M31}  
   \maketitle
\authorrunning{Ma}
\titlerunning{Globular Clusters in M31}

\section{Introduction}

Within the popular $\Lambda$ cold dark matter model for galaxy formation, substructure is expected to be seen in
the outer regions of galaxies as they continue to grow from the accretion and tidal disruption of satellite companions. Globular clusters (GCs), as luminous compact objects that are found out to distant radii in the haloes of massive galaxies, can serve as excellent tracers of substructures in the outer region of their parent galaxy. In addition, clusters which locating at large projected distances from the center of the galaxy would provide the maximum ``leverage'' to constrain the mass distribution of the galaxy \citep[see][and references therein]{Federici93,Evans03,gall05}.
So, detailed studies on GCs in the outer haloes of massive galaxies are very important.

M31, with a distance modulus of 24.47 \citep{Holland98,Stanek98,McConnachie05}, is an ideal
local galaxy for studying GCs, since it is so near, and contains more GCs than all other Local Group galaxies combined \citep{battis87,raci91,harris91,fusi93}. The globular cluster system of M31 has been the subject of deep investigations since the beginning of extragalactic
astronomy \citep{hubble32}. The first catalog of 140 GC candidates in M31 was presented by \citet{hubble32}. From then on, numerous lists of M31 GC candidates were published. The most comprehensive catalog of GC candidates may be the Bologna catalog
\citep{battis80,battis87,battis93}. Bologna catalog contains a total of 827 objects, and all the objects
were classified into five classes by authors' degree of confidence. 353 of these candidates were considered as class A or class B by high level of confidence, and the others fell into class C, D, or E. {\it V} magnitude and $B-V$ color for most candidates were presented in the Bologna catalog. Recent works have searched for fainter GCs in M31 \citep[e.g.,][]{moche98,bh01,kim07}.
In addition, a number of papers by Ma's group have been published on the M31 globular clusters
\citep{fan06,fan08,fan09,jiang03,ma06a, ma06b,ma07a, ma09a, ma09b,ma10,ma11,ma2011,Wang10}, including integrated magnitudes, colors, ages, masses, metallities, and structural parameters.

M31 GC B514 (B for `Bologna', see Battistini 1987) which being called GC 4 in \citet{mackey07}, is the first one of the outer halo cluster known in M31 which locating at a projected distance of $R_p\simeq 55$ kpc. It was detected by \citet{gall05} based on the XSC sources of the All Sky Data Release of the Two Micron All Sky Survey (2MASS, Cutri et al. 2003) within a $\sim 9^\circ \times 9^\circ$
area centered on M31. Now, many new members of M31 halo GC system, which extend to very large radii, have been discovered \citep[e.g.,][]{huxor05,mackey06,mackey07,huxor08,mackey10a}.

In this paper, we will study 10 outer halo GCs of M31 from \citet{mackey06,mackey07} based on 2MASS $JHK_s$ photometry and photometric data in $VI$ bands. In \S 3, we describe the details of our approach to the data reduction. In \S 4, we determine the ages and masses of these outer halo GCs by comparing observational colors with population synthesis models. In \S 5, we present some statistical results for these halo GCs. We discuss the implications of our results and provide a summary in \S 6.

\section{Sample of M31 halo GCs}

The sample of M31 halo globular clusters in this paper is from \citet{mackey06,mackey07}, who estimated their metallicities, distance moduli and reddening values by fitting the Galactic GC fiducials from \citet{brown05} to their observed color-magnitude diagrams (CMDs) in F606W and F814W filter bands of deep images obtained using Advanced Camera for Surveys (ACS) on board the
{\sl Hubble Space Telescope (HST)}. These CMDs extended $\sim 3$ mag below the horizontal branch. In order to study the properties of M31 halo globular clusters, especially,
to determine their ages, their metallisities and reddening values should be determined confidently. In addition, the $V$ and $I$ photometries are needed to construct the color-color diagram (see \S 4 for details).
Till now, there are only these 10 halo globular clusters satisfying these conditions. So, the M31 halo globular clusters which being selected to study here is not
a complete sample. Figure 1 shows the spatial distribution of the 10 halo globular clusters in M31. The large ellipse is the $D_{25}$ boundary of the M31 disk \citep{Vaucouleurs91}.

\begin{figure}
\begin{center}
\includegraphics[height=140mm,angle=-90]{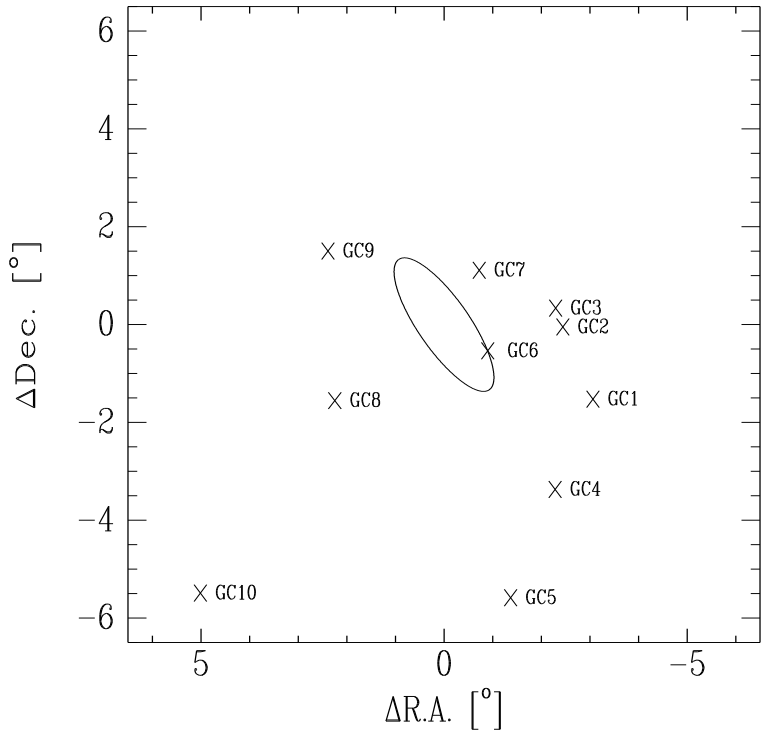}%
\caption{\label{Fig:maps}Spatial distribution of the 10 M31 halo globular clusters. The
ellipse is the $D_{25}$ boundary of the M31 disk \citep{Vaucouleurs91}.}
\end{center}
\end{figure}

\section{Data reductions}

2MASS imaged the entire sky in the near-infrared bands, $J$ ($\rm{1.2~\mu m}$), $H$ ($\rm{1.6~\mu m}$),
and $K_s$ ($\rm{2.2~\mu m}$) down to a limiting sensitivity of 21.6, 20.6, and 20.0 {$\rm mag/arcsec^2$} ($1\sigma$), respectively, with a typical angular resolution of $\sim 2-3''$ (with $1''$ pixels). For these 10 M31 halo GCs, near-infrared images in the $J$, $H$, and $K_s$ bands
were extracted from the 2MASS second data release  \citep{Cutri00} according to the coordinates presented
by \citet{mackey06,mackey07}. In our analysis, the uncompressed atlas images provided by this data release
were used. The pixel size of all of the 2MASS images was $1''$. Exposure time of 1.3 seconds per frame multiplying 6 frames per source makes the total exposure be 7.8 seconds. Figure 2 shows the images of the 10 sample GCs observed in the $J$, $H$, and $K_s$ filters of 2MASS. The image size is $80.0\arcsec\times80.0\arcsec$ for each panel.

\begin{figure}
\begin{center}
\includegraphics[height=220mm,angle=0]{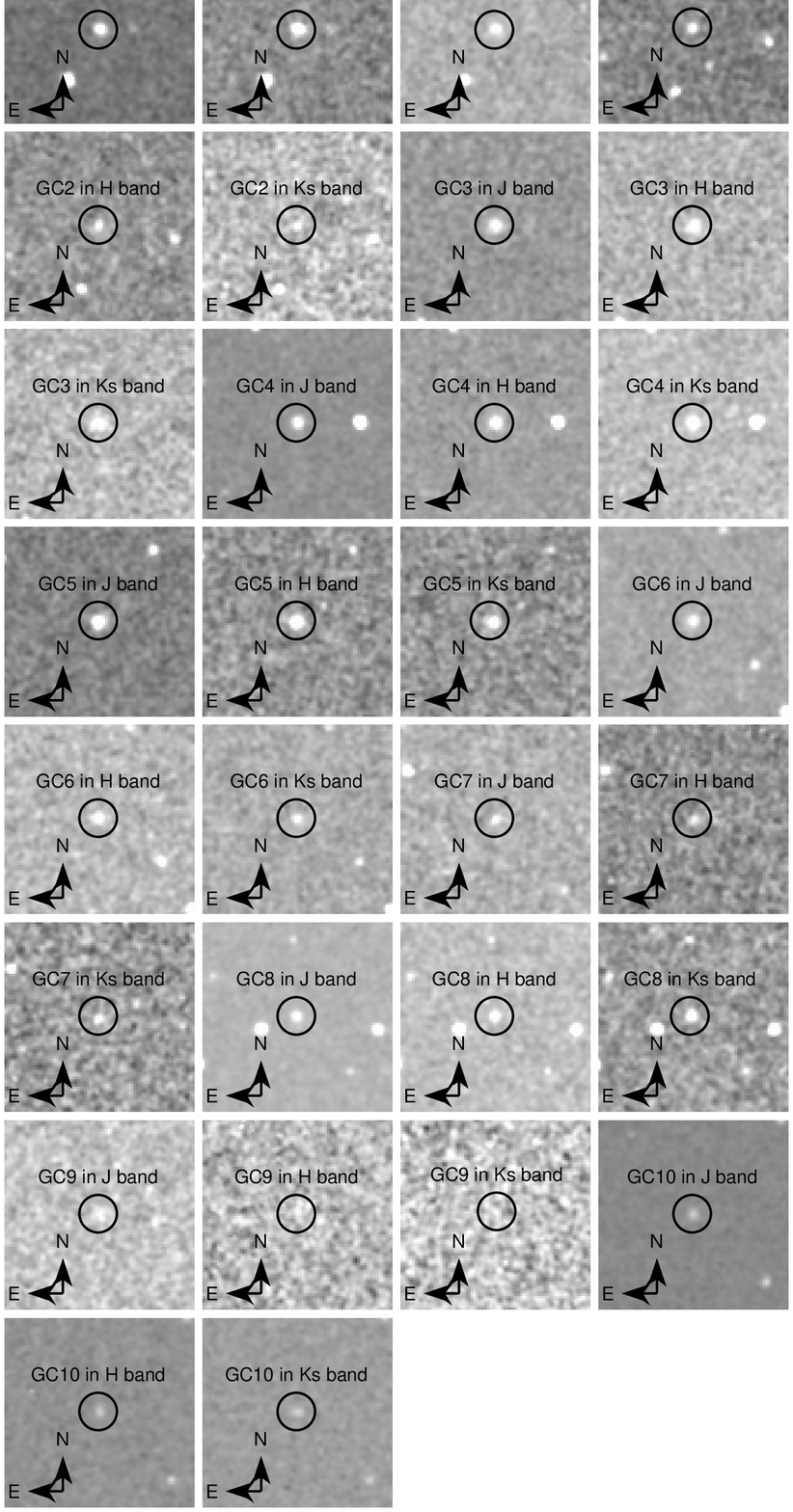}%
\caption{\label{Fig:maps}Images of the 10 sample GCs observed in the $J$, $H$,
and $K_s$ filters of 2MASS. The image size is $80.0\arcsec\times80.0\arcsec$ for each panel.}
\end{center}
\end{figure}

We determined the magnitudes of these GCs using a standard aperture photometry approach, i.e., the
{\sc phot} routine in {\sc daophot} \citep{stet87}. The relevant zero-points for photometry can be
obtained from photometric header keywords of the images. These clusters lie in the halo of M31, and there do not exit extraneous field stars. So, their accurate photometry can be easily derived. We use eight different aperture sizes (with radii of $r_{\rm ap}$ = 3.0, 4.0, 5.0, 6.0, 7.0, 8.0, 9.0, 10.0, 11.0, 12.0, 13.0, and $14.0''$) to ensure that we adopt the most appropriate photometric radius that includes all light from the objects. We
decided on the size of the aperture needed for the photometry based on visual examination. The
proper radius is listed in column (5) of Table 1. For $r_{\rm ap}=9.0''$, we adopted annuli for background
subtraction spanning between 10.0 to $15.0''$. For $r_{\rm ap}=13.0''$, we adopted annuli for background
subtraction spanning between 14.0 to $19.0''$. The calibrated photometry of these objects in $J$, $H$, and $K_s$ filters is summarized in Table 1, in conjunction with the $1\sigma$ magnitude uncertainties obtained from {\sc daophot}.

\section{Cluster ages and masses}

To determine ages for these halo GCs, the colors to use are $(V-I)_0$ and $(J-K_s)_0$, since $(V-I)_0$ varies strongly with age when metallicity being fixed for older
than 1 Gyr (see Figure 3 below). The metallicities of our sample clusters are all determined by \citet{mackey06,mackey07} based on the CMDs constructed using the deep images obtained by the ACS/{\sl HST}. These
metallicities are $-2.14 \le {\rm [Fe/H]}\le -0.70$ which being listed in Table 2, and most clusters are very metal-poor. So, we use the  SSP models of \citet{bc03} (hereafter BC03 models), which have been upgraded from the earlier \citet{bc93,bc96} versions, and now provide the evolution of the spectra and photometric properties for a wider range of stellar metallicities. The BC03 SSP models
based on the Padova-1994 evolutionary tracks include six initial metallicities, $Z= 0.0001, 0.0004, 0.004, 0.008, 0.02\, (Z_\odot)$, and 0.05, corresponding to ${\rm [Fe/H]}=-2.25$, $-1.65$, $-0.64$, $-0.33$, $+0.09$, and $+0.56$. BC03 provide 26 SSP models (both of high and low spectral resolution) using the Padova-1994 evolutionary tracks, half of which were computed based on the
\cite{salp55} IMF with lower and upper-mass cut-offs of $m_{\rm L}=0.1~M_{\odot}$ and $m_{\rm U}=100~M_{\odot}$, respectively. The other thirteen were computed using the \citet{chabrier03} IMF with the same mass cut-offs. In addition, BC03 provide 26 SSP models using the Padova-2000
evolutionary tracks which including six partially different initial metallicities, $Z = 0.0004$, 0.001, 0.004, 0.008, 0.019 $(Z_\odot)$, and 0.03, i.e., ${\rm [Fe/H]}=-1.65, -1.25, -0.64, -0.33, +0.07$, and $+0.29$. In this paper, we will adopt the high-resolution SSP models using the
Padova-1994 evolutionary tracks and a \citet{salp55} IMF to determine the most appropriate ages for these halo GCs since their metallicities are $-2.14 \le {\rm [Fe/H]}\le -0.70$ corresponding to $0.0001 \le {\rm [Fe/H]}\le 0.004$ of  BC03 models. These SSP models contain 221 spectra
describing the spectral evolution of SSPs from $1.0\times10^5$ yr to 20 Gyr. The evolving spectra
include the contribution of the stellar component at wavelengths from 91~\AA~~to $160~\mu$m. BC03 includes the magnitudes in $V$, $I$ and 2MASS bands, making it possible to directly obtain the colors concerned. In Figure 3, we present the $(J-K_s)_0$ versus $(V-I)_0$ color-color diagram for the cluster samples in this paper.
Overplotted with the lines are the colors for the BC03 single generation models for three metallicities, $Z=0.004, Y=0.2400$, $Z=0.0004, Y=0.2310$, and $Z=0.0001, Y=0.2303$. The age range is from 1 Gyr to 20 Gyr. The colors of ($V-I$) are from \citet{huxor08} except for GC6, of which the color of ($V-I$) is from the RBC v.4.0 \citep{gall04,gall06,gall07}, since \citet{huxor08} did not present it. The error bars of $(V-I)$ are the photometric
uncertainties, which are estimated to be $\sim 0.03$ mag as \citet{huxor08} presented except for GC6. For GC6,
which is also called B298, the magnitudes in $V$ and $I$ bands are 16.59 and 15.46, respectively. The photometric uncertainties are adopted to be 0.05 as \citet{gall04} suggested. The integrated cluster colors have been dereddened by the $E(B-V)$ values given by \citet{mackey07}. These reddening values are listed in Table 2. The interstellar extinction curve, $A_{\lambda}$, is taken from \citet{car89}, $R_V=A_V/E(B-V)=3.1$. From Figure 3, we can see that, the $V-I$ color is very sensitive to age when metallicity is fixed. $J-K_s$ and
$V-I$ colors of each object were compared with a model of $Z$ that is most close to the metallicity obtained by \citet{mackey07} to determine ages for clusters. The ages obtained in this paper are listed in Table 2, which all are older than 3~Gyr. The 4 clusters are older than 10~Gyrs,
and the other 6 are in intermediate ages between $3-8$~Gyrs

\begin{figure}
\begin{center}
\includegraphics[height=140mm,angle=-90]{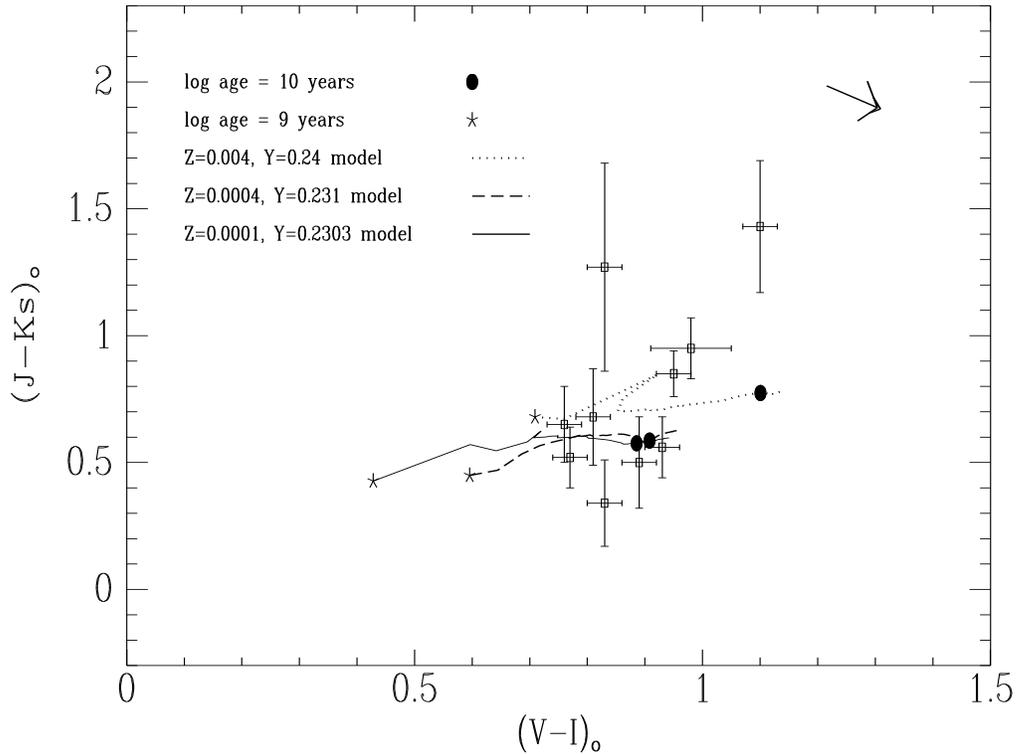}%
\caption{\label{Fig:maps}$(V-I)_0$ vs. $(J-Ks)_0$ for M31 halo GCs. The photometric errors are represented by the error bars. The clusters have been dereddened by the reddening values from \citet{mackey07}. BC03 models for three separate $Z$ values are shown. The arrow represents the direction of the reddening vector.}
\end{center}
\end{figure}

Cluster masses can be obtained by comparing the measured luminosity in the $V$ band with the theoretical mass-to-light ($M/L$) ratios. The $M/L$ ratios are a function of the cluster age and metallicity. The
mass-to-light ratios of these GCs, calculated based on the metallitity adopted and the age obtained in this paper, are listed in Table 3 for the BC03 SSP models. The magnitudes in $V$ band are from \citet{huxor08}, and from the RBC v.4.0 \citep{gall04,gall06,gall07} (only for GC6). Distance moduli for these GCs are from \citet{mackey07}. The derived masses are listed in Table 3, from which we can see that the masses of these outer halo GCs are from $7.0\times 10^4 M_{\odot}$ to $1.02\times 10^6~M_{\odot}$.

\section{Some statistical properties of sample halo globular clusters}

\subsection{Radially-dependent properties of sample halo globular clusters}

\begin{figure}
\begin{center}
\includegraphics[height=140mm,angle=-90]{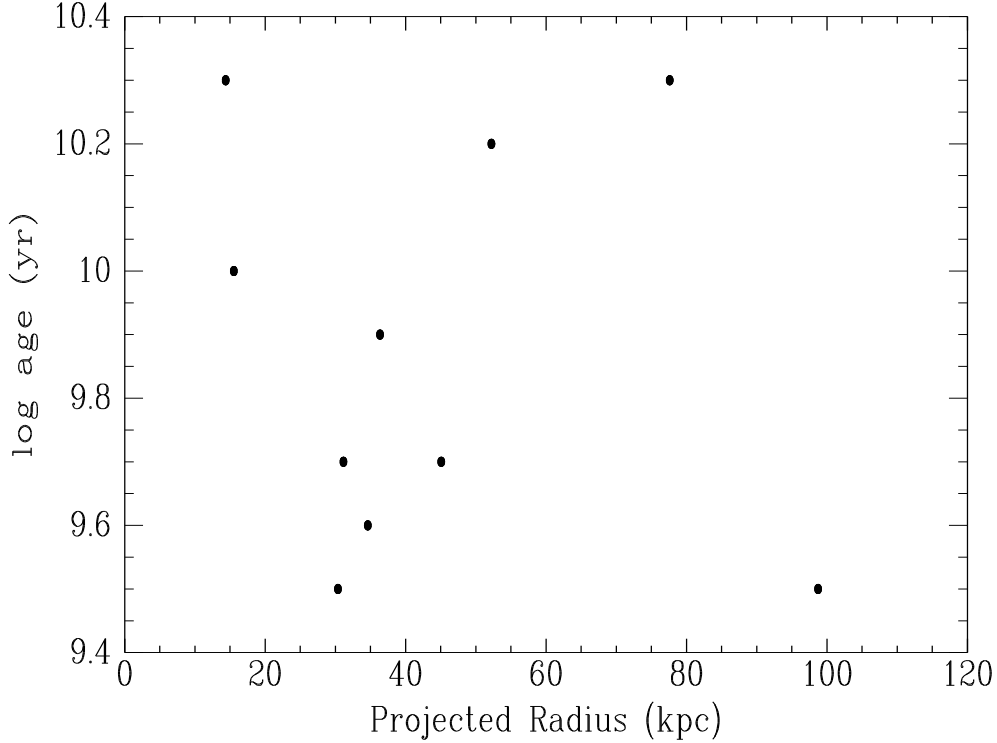}%
\caption{\label{Fig:maps}Age as a function of galactocentric distance for these 10 outer halo GCs in M31.}
\end{center}
\end{figure}

Galactocentric coordinates for the sample halo globular clusters were computed relative to an adopted M31 central position. In this paper, we adopted a central position for M31 at $\rm \alpha_0=00^h42^m44^s.30$ and
$\rm \delta_0=+41^o16'09''.0$ (J2000.0) following \citet{huchra91} and \citet{perr02}.
Formally,

\begin{equation}
X=A\sin\theta+B\cos\theta \quad ;
\end{equation}

\begin{equation}
Y=-A\cos\theta+B\sin\theta \quad ,
\end{equation}
where $A=\sin(\alpha-\alpha_0)\cos\delta$ and $B=\sin\delta \cos\delta_0 - \cos(\alpha-\alpha_0)
\cos\delta \sin\delta_0$. We adopt a position angle of $\theta=38^\circ$ for the major axis of M31 \citep{Kent89}. The $X$ coordinate is the position along the major axis of M31, where positive $X$ is in the northeastern direction, while the $Y$ coordinate is along the minor axis of the M31 disk, increasing towards the northwest. Projected radius is $\sqrt{X^2+Y^2}$.

\subsubsection{Age as a function of projected distance}

\citet{Gratton85} derived the ages of 26 Galactic GCs using the data on the main sequence photometry available, and found that as the projected galactocentric distance increases the ages of the clusters decrease. However, this relationship in Gratton's analysis is based on the photometry for the four most distant clusters, the quality of which may be not high \citep[see][for details]{SC89}. \citet{SC89} calculated ages for 32 Galactic GCs using the magnitude difference between the horizontal branch (HB) and the main sequence turnoff (MSTO). The statistical result showed that there do not exist a relation between age and galactocentric distance. Subsequent work by \citet{Chaboye92}, \citet{Sarajedini95},
and \citet{Chaboye96} reinforced the results of \citet{SC89} using larger and more reliable data sets of GC ages as well as a variety of techniques for measuring them. Based on the uniform photometric data set provided by the ACS/{\sl HST} Galactic GC Treasury project \citep{Sarajedini07}, \citet{MF09} and \citet{Dotter10}
determine ages from a large sample of GCs. The statistical results of \citet{MF09} and \citet{Dotter10}
also showed that there is not any relation between age and galactocentric distance. Figure 4 shows the distribution of ages against projected galactocentric distance for these
10 outer halo GCs. Here we note that no relationship exists. This conclusion is consistent with \citet{SC89} and their series works.

\subsubsection{Mass as a function of projected distance}

Figure 5 shows the distribution of masses against projected galactocentric distance for these 10 outer halo GCs. However, there is not any relationship between mass and galactocentric distance.

\begin{figure}
\begin{center}
\includegraphics[height=140mm,angle=-90]{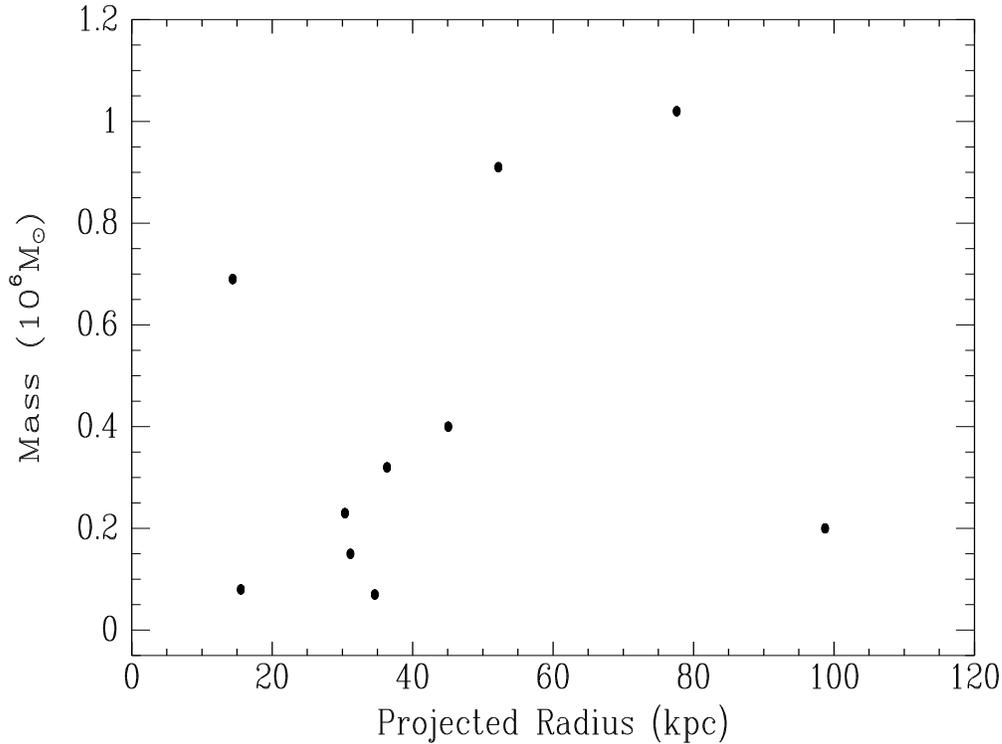}%
\caption{\label{Fig:maps}Mass as a function of galactocentric distance for these 10 outer halo GCs in M31.}
\end{center}
\end{figure}

\subsubsection{Metallicity as a function of projected distance}

One early formation model by \citet{eggen62} argues for a single, large-scale collapse of material to form galactic bodies such as the Milky Way, in which the enrichment
timescale is shorter than the collapse time, the halo stars and globular clusters should show large-scale metallicity gradients; a principal competing model maintains that formation occurred through random mergers of fragmented gas clouds over the course of billions of years, implying a hierarchical origin \citep{SZ78}, so there should be a homogeneous metallicity distribution.

For M31 globular clusters, there are some inconsistent conclusions, for example, \citet{vand69} showed that there is little or no evidence for a correlation
between metallicity and projected radius, but most of his
clusters were inside $50''$; however, some authors \citep[see, e.g.][]{hsv82, sha88, hbk91} have
presented evidence for a weak but measurable metallicity gradient as a function of projected radius. \citet{bh00}, \citet{perr02} and \citet{fan08} confirmed the latter result based on their large sample of spectral
metallicity and color-derived metallicity.

Figure 6 shows the distribution of metallicities against projected galactocentric distance for these 10 outer halo GCs. Clearly, the dominant feature of this diagram is the
scatter in metallicity at any radius. However, the most distant globular cluster (GC10) is really very metal-poor. We also noted that, the metallicity of GC6 is very
low (${\rm [Fe/H]}=-2.14$), and the metallicity of GC7
is very high (${\rm [Fe/H]}=-0.7$). However, these two clusters lie nearly the same distance from the center of M31 (see Figure 6). This result is consistent with
\citet{Yang10} finding for the halo populations of M33:
the halo populations show no trend of metal abundance with radial position.

\begin{figure}
\begin{center}
\includegraphics[height=140mm,angle=-90]{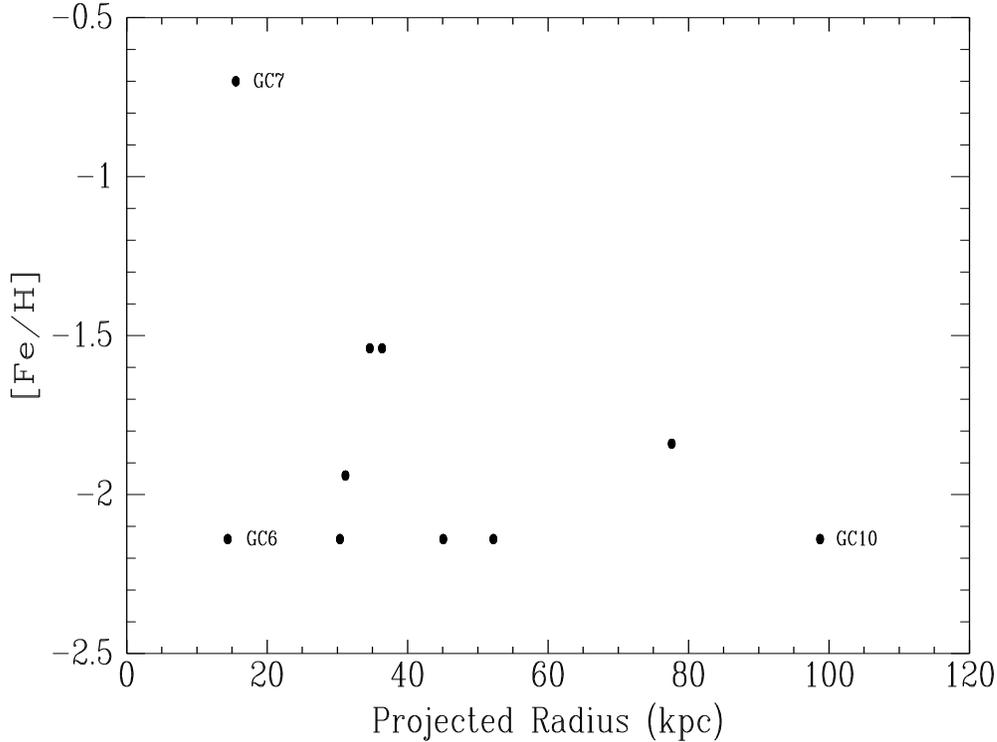}%
\caption{\label{Fig:maps}Metallicity as a function of galactocentric distance for these 10 outer halo GCs in M31.}
\end{center}
\end{figure}

\subsection{Relationship between age and metallicity for sample halo globular clusters}

The Galactic GC system is well known to obey a clear age-metallicity relationship, in the sense of older GCs being characterized by generally lower metallicities
\citep[e.g.,][]{SC89,Chaboye96}. \citet{fan06} showed that although there is significant scatter in metallicity at any age, there is a notable lack of young metal-poor and
old metal-rich GCs, which might be indicative of an underlying age-metallicity relationship among the M31 GC population. Figure 7 shows the metallicity as a function of age for the outer halo GC sample discussed in this paper. It is evident that there are not any relationship between age and metallicity.

\begin{figure}
\begin{center}
\includegraphics[height=140mm,angle=-90]{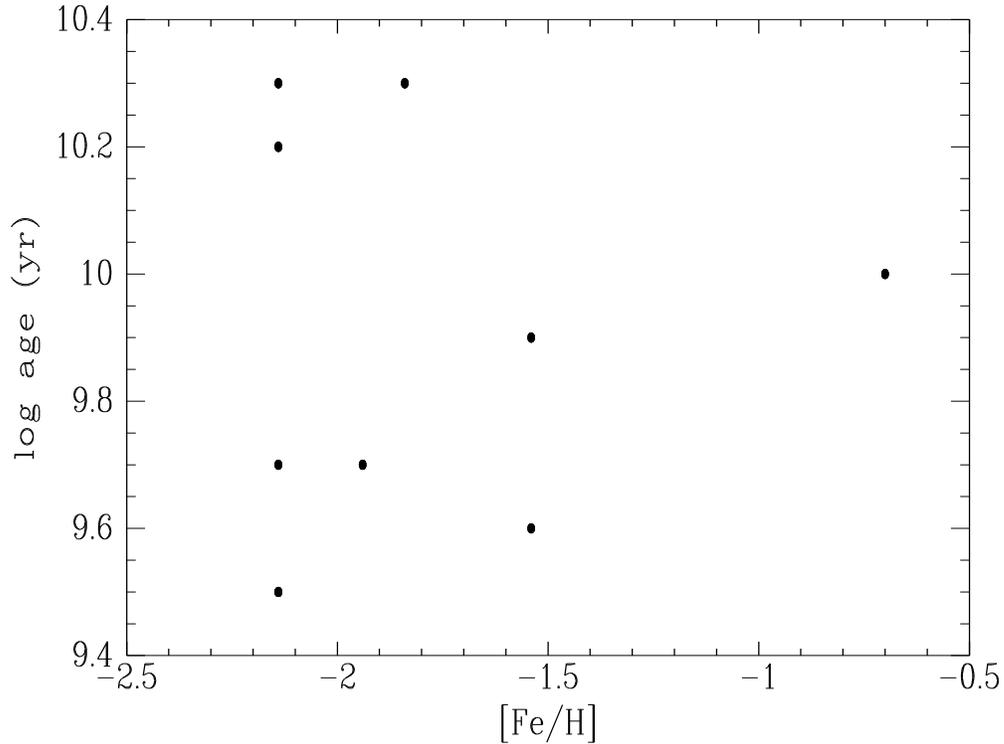}%
\caption{\label{Fig:maps}Relationship between age and metallicity for these 10 outer halo GCs in M31.}
\end{center}
\end{figure}

\subsection{Relationship between age and mass for sample halo globular clusters}

Figure 8 shows the mass as a function of age for the outer halo GC sample discussed in this paper. It appears that a general trend between GC age and mass may exist, in
the sense that the more massive clusters are older except for GC7, which has the mass of $8.0\times 10^4 M_{\odot}$. However, its age is $\sim 10$~Gyr.

\begin{figure}
\begin{center}
\includegraphics[height=140mm,angle=-90]{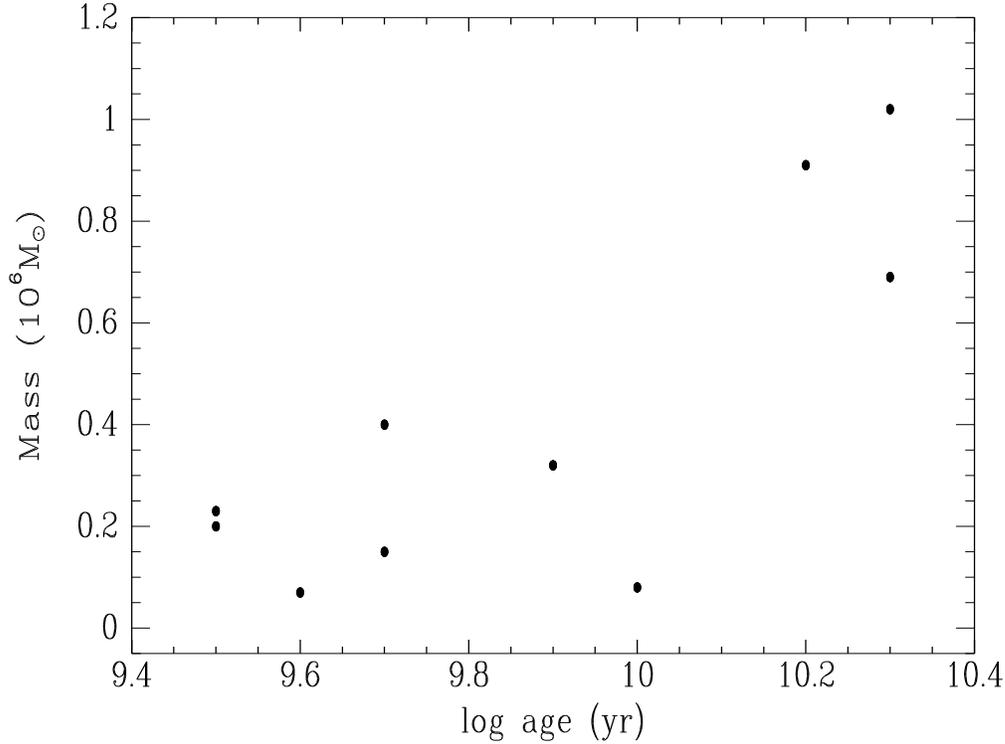}%
\caption{\label{Fig:maps}Relationship between age and mass for these 10 outer halo GCs in M31.}
\end{center}
\end{figure}

\section{Summary and discussion}

This paper firstly presented photometric results in $J$, $H$, and $K_s$ from 2MASS imaging for 10 classical globular clusters in the far outer regions of M31. Combined with the $V$ and $I$ photometric data from previous literature, we constructed the color-color diagram between $J-K_s$ and $V-I$. By comparing the integrated photometric measurements with evolutionary models, we estimate the ages of these clusters. The results showed that, all of these clusters are older than 3~Gyrs. In addition, we estimated the masses for these outer halo GCs which being from $7.0\times 10^4 M_{\odot}$ to $1.02\times 10^6~M_{\odot}$.

In their seminal work, \citet{SZ78} examined the abundances of a sample of Galactic GCs, and found that there is no radial abundance gradient in the cluster system of the outer halo. They argued that this fact, along with a broad spread in the color distribution on the horizontal
branch for distant GCs, implied that a significant fraction of GCs at Galactocentric ${\rm {radii\ge 8~kpc}}$ formed in smaller ``proto-galactic fragments'' that were subsequently accreted into the Galactic potential well. Recent work by \citet{MF09}, \citet{Gratton10}, and \citet{Dotter10} showed that the Galactic globular clusters
(with Galactocentric ${\rm {radii > 8~kpc}}$) exhibit properties consistent with having been accreted in a relatively slow and chaotic fashion from dwarf satellite galaxies of the Milky Way \citep[see also][]{Yang10}. Based on a sample of newly discovered GCs from the Pan-Andromeda Archaeological Survey (PAndAS) in combination with previously cataloged objects, \citet{mackey10b} mapped the spatial distribution of GCs in the M31 halo. The GCs studied here are also included in the sample of \citet{mackey10b}. \citet{mackey10b} found that, at projected radii beyond ${\rm 30~kpc}$, where large coherent stellar streams are readily distinguished in the field, there is a striking correlation between these features and the positions of the GCs. Using a simple Monte Carlo simulation, they computed the probability that it could be due to the chance alignment of GCs smoothly distributed in the M31 halo, and found the likelihood of this possibility being low, below 1\%, which implying that the majority of the remote GC system of M31 were accreted from satellite galaxies.

It is interesting that GC2 and GC3, the ages and metallicities of which are nearly the same (The ages for GC2 and GC3 are $\rm {\log age = 9.7}$ and 9.5, respectively, and the metallicities for GC2 and GC3 are $\rm {[Fe/H] = -1.94}$ and $-2.14$, respectively.).
In addition, from \citet{mackey07}, the distance moduli of GC1 and GC2 are $(m-M)_0=24.32\pm0.14$ and $(m-M)_0=24.37\pm0.15$, i.e. they are very close. So, we argued that GC2 and GC3 were accreted from the same satellite galaxy, if they did not form in M31.

\newpage
\begin{acknowledgements}
We would like to thank the referee, Professor Jingyao Hu, for providing rapid and thoughtful report that helped improve the original manuscript greatly. This work was supported by the Chinese National Natural Science Foundation grands No. 10873016, and 10633020, and by National Basic Research Program of China (973 Program), No. 2007CB815403.
\end{acknowledgements}

\newpage
\begin{table}
\begin{center}
\renewcommand\arraystretch{1}
\caption{Photometry in 2 MASS $J, H$ and $K_s$ for 10 halo globular clusters in M31.}
\begin{tabular}{ccccc}
\hline\hline
 Identifier  &  $J$  &   $H$   &   $K_s$ & $r_{\rm ap}$ \\
             &  (mag)&   (mag) &    (mag)&   ($''$)     \\
\hline
 GC1 & $14.116\pm0.061$ & $13.497\pm0.062$ & $13.729\pm0.158$ & 13.0\\
 GC2 & $15.284\pm0.106$ & $14.419\pm0.096$ & $14.564\pm0.156$ & 9.0 \\
 GC3 & $14.486\pm0.071$ & $14.404\pm0.101$ & $13.904\pm0.093$ & 9.0 \\
 GC4 & $14.229\pm0.065$ & $13.318\pm0.055$ & $13.627\pm0.098$ & 13.0\\
 GC5 & $14.314\pm0.058$ & $13.947\pm0.070$ & $13.425\pm0.069$ & 9.0 \\
 GC6 & $14.797\pm0.076$ & $14.495\pm0.103$ & $13.800\pm0.090$ & 13.0\\
 GC7 & $16.322\pm0.158$ & $15.685\pm0.199$ & $14.861\pm0.138$ & 9.0 \\
 GC8 & $14.741\pm0.084$ & $13.941\pm0.111$ & $14.199\pm0.155$ & 13.0\\
 GC9 & $16.723\pm0.308$ & $15.293\pm0.199$ & $15.373\pm0.277$ & 9.0 \\
 GC10& $14.583\pm0.076$ & $14.268\pm0.121$ & $13.887\pm0.124$ & 13.0\\
\hline
\end{tabular}
\end{center}
\end{table}

\begin{table}
\begin{center}
\renewcommand\arraystretch{1}
\caption{Age estimates for 10 halo globular clusters in M31.}
\begin{tabular}{cccccccc}
\hline\hline
 Identifier  &  $V-I$  &   $J-Ks$   & $E(B-V)$&  $(V-I)_0$  &  $(J-Ks)_0$  &  Metallicity & Age     \\
             &  (mag)  &   (mag)    &    (mag)&   (mag)     &    (mag)     &     (dex)    & [log yr]\\
\hline
 GC1 & $0.98\pm0.03$ & $0.387\pm0.17$ & $0.09\pm0.01$ & $0.83\pm0.03$ & $0.34\pm0.17$ & $-2.14$ & $9.7\pm0.1$ \\
 GC2 & $0.94\pm0.03$ & $0.720\pm0.19$ & $0.08\pm0.01$ & $0.81\pm0.03$ & $0.68\pm0.19$ & $-1.94$ & $9.7\pm0.1$ \\
 GC3 & $0.95\pm0.03$ & $0.582\pm0.12$ & $0.11\pm0.01$ & $0.77\pm0.03$ & $0.52\pm0.12$ & $-2.14$ & $9.5\pm0.2$ \\
 GC4 & $1.08\pm0.03$ & $0.602\pm0.12$ & $0.09\pm0.01$ & $0.93\pm0.03$ & $0.56\pm0.12$ & $-2.14$ & $10.2\pm0.1$\\
 GC5 & $1.08\pm0.03$ & $0.889\pm0.09$ & $0.08\pm0.01$ & $0.95\pm0.03$ & $0.85\pm0.09$ & $-1.84$ & $10.3\pm0.1$\\
 GC6 & $1.13\pm0.03$ & $0.997\pm0.12$ & $0.09\pm0.01$ & $0.98\pm0.03$ & $0.95\pm0.12$ & $-2.14$ & $10.3\pm0.1$\\
 GC7 & $1.20\pm0.03$ & $1.461\pm0.21$ & $0.06\pm0.01$ & $1.10\pm0.03$ & $1.43\pm0.21$ & $-0.70$ & $10.0\pm0.1$\\
 GC8 & $1.04\pm0.03$ & $0.542\pm0.18$ & $0.09\pm0.01$ & $0.89\pm0.03$ & $0.50\pm0.18$ & $-1.54$ & $9.9\pm0.2$ \\
 GC9 & $1.07\pm0.03$ & $1.350\pm0.41$ & $0.15\pm0.01$ & $0.83\pm0.03$ & $1.27\pm0.41$ & $-1.54$ & $9.6\pm0.1$ \\
 GC10& $0.91\pm0.03$ & $0.696\pm0.15$ & $0.09\pm0.01$ & $0.76\pm0.03$ & $0.65\pm0.15$ & $-2.14$ & $9.5\pm0.1$ \\
\hline
\end{tabular}
\end{center}
\end{table}

\begin{table}
\begin{center}
\renewcommand\arraystretch{1}
\caption{Mass estimates for 10 halo globular clusters in M31.}
\begin{tabular}{ccccc}
\hline\hline
 Identifier  &  $V$  &   $(m-M)_0$   &  $M/L_V$          &  Mass           \\
             & (mag) &   (mag)       &  $(M/L_V)_\odot$  & $(10^6M_\odot)$ \\
\hline
 GC1 & 16.05 & 24.41 & 1.63 & $0.40\pm0.07$\\
 GC2 & 16.98 & 24.32 & 1.63 & $0.15\pm0.03$\\
 GC3 & 16.31 & 24.37 & 1.17 & $0.23\pm0.04$\\
 GC4 & 15.76 & 24.35 & 3.01 & $0.91\pm0.17$\\
 GC5 & 16.09 & 24.45 & 4.30 & $1.02\pm0.19$\\
 GC6 & 16.59 & 24.49 & 4.30 & $0.69\pm0.32$\\
 GC7 & 18.27 & 24.13 & 3.51 & $0.08\pm0.02$\\
 GC8 & 16.72 & 24.43 & 2.40 & $0.32\pm0.06$\\
 GC9 & 17.78 & 24.22 & 1.50 & $0.07\pm0.01$\\
 GC10& 16.50 & 24.42 & 1.21 & $0.20\pm0.04$\\
\hline
\end{tabular}
\end{center}
\end{table}

\end{document}